\documentstyle[eqsecnum,aps,pre,psfig]{revtex}
\begin{document}
\twocolumn
%
\wideabs{
\title{Long-range fractal correlations in literary corpora}

\author{Marcelo A. Montemurro~\cite{email1}
and Pedro A. Pury~\cite{email2}}

\address{Facultad de Matem\'atica, Astronom\'\i a y F\'\i sica \\
Universidad Nacional de C\'ordoba \\
Ciudad Universitaria, 5000 C\'ordoba, Argentina}

\date{\today}

\maketitle

\begin{abstract}
In this paper we analyse the fractal structure of long human-language
records by mapping large samples of texts onto time series.
The particular mapping set up in this work is inspired on
linguistic basis in the sense that is retains {\em the word} as the
fundamental unit of communication. The results confirm that beyond
the short-range correlations resulting from syntactic rules acting
at sentence level, long-range structures emerge in large written
language samples that give rise to long-range correlations in
the use of words.
\end{abstract}

\vspace{0.5cm}
}
\narrowtext

\section{Introduction}
\label{sec:Intro}

The human language faculty reveals as a phenomenon of remarkable
complexity, intimately interwoven with all our superior mental
functions~\cite{Pink}. Its role in the modelling and
categorisation of factual experience by the mind cannot be
downplayed. Moreover, it has been suggested that the outburst of
symbolic and abstract imagery pervading the archeological
records signalling the transition between the middle and upper
Paleolithic periods, might be interpreted as the hallmark of an 
associated, and rather sudden, transition in the linguistic
performance of modern humans~\cite{Mell98,TM00}. 
The evolutionary advantage of complex syntactic communication stems
from the capacity that it confers to language of coding complex
information using moderate symbolic resources~\cite{NPJ00}. 
In the traditional areas of linguistics much success has been
attained in dissecting language structure from sentence level 
down the morphological rules of word formation. 
However, effective linguistic communication is a complex cognitive
process~\cite{ADFH92} where sequences of sentences cohere in the
assemblage of larger meaningful structures.
At this point the study of language becomes susceptible of being
complemented by using techniques that have been successfully 
applied on other complex systems in nature that also allow a 
variety of interdisciplinary approaches.

In particular, one question that still remains open to discussion 
is the ultimate origin of long-range correlations in complex
information systems like the genetic code or human language.
The main goal of this work is to examine the existence of long-range
correlations in the use of words in written language drawing on
methods from statistical time series analysis. To that end, we first
describe the particular mapping scheme used to translate the given
sequence of words present in a text sample into a time series. Then,
we devote a section to explain the {\it rescaled range} analysis and
how it can be used to infer the presence of long-range correlations
in time series. Finally, the results of our systematic analysis are
presented followed by concluding remarks.

\section{The mapping of texts onto time series}
\label{sec:Zipf}

The statistical analysis of long symbolic structures had an
accelerated development right after the first complete DNA sequences
started to become available, around a decade ago~\cite{DNA}.
Basically, the original techniques tried to create a random walk out
of the particular symbolic sequence by mapping each distinct symbol
onto a step in an independent direction in space.
Thus, in a genetic sequence, made up of the concatenation of four
independent nucleotides, represented by four letters, each instance
of these symbols was translated onto an elementary jump along
independent directions in a four dimensional space.
Although different alternatives for the particular details of the
mappings were put forward, all were devised to keep the original
sequence's structure at symbol level translated into the time series.
Then, by applying a set of standard statistical techniques, such as
power spectrum analysis, detrended fluctuation analysis or rescaled
range analysis among others, it was possible to unravel the existence
of long-range correlations in biosequences~\cite{DNA,Voss94}.

Written samples of natural languages, are also complex information
carriers to be read sequentially. Thus, similar techniques as
those employed in genetic sequence analysis were adapted with
appropriate modifications~\cite{SZZ93,ASE+94,Ebe}.
The different procedures shared one common point, namely, that
the mappings were invariably performed at letter level.
However, it has never been discussed in the literature so far 
whether the mapping at letter level is indeed adequate in the 
case of human language. It is crucial to note that the specific
coding of words in some particular spelling or phonetic system is
alien to the linguistic structure of communication, as is clear by
noting that a given text can be readily coded in sign languages, 
for instance. Therefore, a better mapping, founded on linguistics
basis, should recognise the symbols at the minimum level intrinsic 
to the communication process.
In what follows we propose the simplest step in that direction,
which consists in taking word tokens as the fundamental units of
communication. By doing this we can assert that the mapped sequences
do not carry any structure below word level.

In order to make clear the details of the actual translation of a
given text onto a time series, a brief detour around Zipf's analysis
is required~\cite{Zipf}. It basically consists in counting the number
of occurrences of each distinct word in a given text sample, and then
producing a list of all these words ordered by decreasing frequency.
Each word in this list can be identified by an index equal to its
rank, $r$, in the list, that is,  the most frequent word has index
$r=1$, the second in the list is given $r=2$, and so on.  Words that
occur the same number of times are ordered arbitrarily in their
corresponding rank interval.

By means of Zipf's analysis each word in the original text can be
replaced by its corresponding index $r$. Then, at position $t$
starting from the beginning of the text, we have the corresponding
index $r(t)$. According to this mapping, the whole text becomes a
time series of ranks, namely $\{r(t)\}_{t=1}^{T}$, where $T$ stands
as the total length of the text. This particular numerical assignment
may seem arbitrary at first, and in fact a different choice would
certainly work as well. However, the selection of the rank as an
indexing key to the words may be rendered more natural if we think of
it as the  assignment that minimises the effort in lexical access in
the rank-ordered list of words in writing the whole text. When a word
is required at position $t$ in the text, then it must be picked from
position $r(t)$ in the list of words ordered by decreasing frequency.

In order to compare different time series generated by this mapping,
it is useful to rescale the series to zero mean and unit variance.
Thence, we define the standard deviation of the rank sample values 
as follows
$\hat\sigma = \sqrt{\left< r(t)^2 \right> - \left< r(t) \right>^2}$,
where the symbol $\left< \dots \right>$ denotes an arithmetic 
average over the whole series of ranks.
Then, the quantity $\xi(t)$ is defined as
\begin{equation}
\label{xi}
\xi(t)=\frac{r(t)- \left< r(t) \right>}{\hat\sigma} \;\;.
\end{equation}
In this way, the sequence of normalised increments $\xi(t)$ can now
be regarded as the sequence of elementary jumps in a random walk each
taken at a discrete time $t$.

\section{Rescaled Range Analysis}
\label{sec:RRS}

Hydrology is the oldest discipline in which the presence 
of noncyclic very long-term dependence has been reported.  
Particularly, the {\em rescaled range} analysis was 
introduced by Harold E.\ Hurst~\cite{Hurst} when he 
was studying the Nile in order to describe the long-term 
dependence of  water levels in the river and reservoirs.
Later, this statistical technique was further developed and 
applied by Mandelbrot and Wallis~\cite{MW69b,MW69d,MW69e}.

Let $\xi(t)$, $1 \leq t \leq T$, be the normalised 
sequence of increments of a process in discrete time.
In our case, $\xi(t)$ is the normalised ordered sequence of 
ranks from a text corpus of $T$ words, as described in the
previuos section.
From this sequence, the record
\begin{equation}
X(t) = \sum_{u=1}^t \xi(u)
\end{equation}
is constructed.
Viewing the $\xi(t)$ as spatial increments in a one-dimensional
discrete random walk, $X(t)$ is the position of the walker from 
the starting point at time $t$.
For any given integer span $s > 1$ and any initial time $t$,
a detrended subrecord $D(u,t,s)$, for $0 \leq u \leq s$,
can be defined as
\begin{equation}
D(u,t,s) = X(t + u) - X(t) - \frac{u}{s} 
\,\left( X(t + s) - X(t) \right) \;\;.
\label{D}
\end{equation}
In this quantity, the mean $\sum_{w=1}^{s} \xi(t + w) / s$
was substracted to remove the trend in the subrecord.
The cumulated range $R(t,s)$ of the subrecord is defined by
\begin{equation}
R(t,s) = 
\max_{0 \leq u \leq s} D(u,t,s) -
\min_{0 \leq u \leq s} D(u,t,s)
\label{R}
\end{equation}
and the variance $S^2(t,s)$ of the subrecord is defined by
\begin{equation}
S^2(t,s) = \frac{1}{s} \sum_{w=1}^{s} \xi^2(t + w) -
\left( \frac{1}{s}\sum_{w=1}^{s} \xi(t + w) \right)^2 \;\;.
\label{S2}
\end{equation}

For many time series of natural phenomena the average of the
sample values of $R(t,s)/S(t,s)$, carried over all admissible 
starting points $t$ within the sample, follows the Hurst's law:
${\cal E}[R(t,s)/S(t,s)] \sim s^H$ with $H > 1/2$.
Hurst's observation is remarkable considering the fact that in 
the {\em absence} of long-run statistical dependence one should
find $H = 1/2$, for processes with finite variance.
For example, for a stationary Gaussian process $\xi(t)$ with
$\left< \xi(t) \right> = 0$ and $\left< \xi^2(t) \right> = 1$, 
Feller~\cite{Feller} has analytically proved that
\begin{equation}
\lim_{s \rightarrow \infty} s^{-1/2} {\cal E}[R(t,s)/S(t,s)] =
\sqrt{\frac{\pi}{2}} \;\;.
\label{gaussian}
\end{equation}
Additionally, Mandelbrot and Wallis~\cite{MW69e}  showed that the
$s^{1/2}$~law also applies to processes of independent increments
having a variety of distributions: truncated Gaussian, hyperbolic,
and (highly skewed) log-normal. Moreover, they also showed that when
the increments of the process are statistical dependent but the
dependence is limited to the short run, the $s^{1/2}$~law holds
asymptotically.
The effect of strong cyclic components was also studied. 
When a white Gaussian noise (of zero mean and unit variance) 
is superimposed with a purely periodic process of amplitude $A$, 
it can be seen that
\begin{equation}
\lim_{s \rightarrow \infty} s^{-1/2} {\cal E}[R(t,s)/S(t,s)] =
\sqrt{\frac{\pi}{2}} \left(1 + \frac{A}{2} \right)^{-1/2} \;\;.
\label{Feller}
\end{equation}
The value of the limit as well as the speed with which this 
limit is attained is dependent on $A$. When $A$ increases, 
${\cal E}[R(t,s)/S(t,s)]$ takes an even longer time to reach 
the $s^{1/2}$~law. Moreover, the transition to the asymptote 
is non monotonic, but typically exhibits a series of oscillations.

The above comments support the key idea that $s^{1/2}$~law holds for
every process for which long-term dependence is unquestionably absent
and does not hold for processes exhibiting noncyclic
long-term statistical dependence. Thus, a Hurst exponent of $H = 1/2$
corresponds to the vanishing of correlations between past and future
spatial increments in the record. For $H > 1/2$ one has persistent
behaviour, which means a positive increment in the past will on the
average lead to a positive increment in the future.
Conversely a decreasing trend in the past implies on the average a
sustained decrease in the future. Correspondingly, the case $H < 1/2$
denotes antipersistent behaviour.

In all the above discussion the focus was on the scaling properties
of the ratio $R(t,s)/S(t,s)$. It is important to note that the
introduction of the denominator $S(t,s)$ becomes ever more relevant
for processes that deviate from the Gaussian and/or have long-term 
dependence~\cite{MW69e}.
Futhermore, the ratio $R(t,s)/S(t,s)$ has a better sampling 
stability in the sense that the relative deviation, defined by
$\sqrt{Var[R(t,s)/S(t,s)]}/{\cal E}[R(t,s)/S(t,s)]$, is smaller
than any alternative expression used to study long-term dependence.

\section{Estimation of $H$ in literary corpora}
\label{sec:H}

In this section we shall explain the implementation of the {\em
rescaled range} analysis, as well as present the results obtained
after performing the experiments on texts coded as a normalised
sequences of ranks.

As it was stated in the previous section, the method is based on 
the estimation of sample averages of the ratio $R(t,s)/S(t,s)$, 
for proper choice of different values of the integer span $s$.
We shall follow the nomenclature and prescriptions introduced by
Mandelbrot and Wallis~\cite{MW69d} in the construction of the
R/S diagrams from the experimental data.
We first compute the exponent $m$ such that $2^m \leq T < 2^{m+1}$,
and then we select the value of the span $s$ from the decreasing
sequence of integers: 
$\{ T/2^{p}, \,p = 0, 1, \dots, m-2 \}$.
For each $s$, we choose the starting points:
$t = q\,s+1, \,q = 0, 1, \dots, 2^p-1$
in order to construct $2^p$ nonoverlapping
detrended subrecords $D(u,t,s)$ from the Eq.~(\ref{D}).
Thus, for a given value of the index $p$, we have specified a
value of $\log s$ which is marked on the axis of abscissas.
In this way, we can reckon $2^p$ values of $\log[R(t,s)/S(t,s)]$
corresponding to the different starting points $t$, which
are plotted as ordinates (marked by + signs). 
For each $s$, the logarithm of the sample average of the quantities 
$R(t,s)/S(t,s)$ over the different starting points is also marked 
in the R/S diagram.
Then, the value of $H$ is estimated by linear regression of
$\log{\cal E}[R(t,s)/S(t,s)]$ vs.\ $\log s$.
The slope is calculated using the linear least-square method
and the error is evaluated taking the uncertainty in the ordinates,
associated with each value of the abscissa ($\log s$), equal to one 
quarter of the amplitude between the corresponding extreme points.

There are two pitfalls related to the R/S diagrams. First, for small
values of $\log s$ (short subrecords) there is large scattering in 
the values of $\log[R(t,s)/S(t,s)]$. Second, for large values of
$\log s$ we have very few nonoverlapping subrecords, thus narrowing 
the R/S diagram. This tightening involves a deceiving evaluation of
uncertainty in the ordinate.
Hence, when fitting a straight trendline, small and large
values of $\log s$ should be neglected.
Given a value of $p$ we have $2^p$ subsamples each of which 
have $T/2^p \geq 2^{m-p}$ ranks.
Therefore, we will retain the values of the span $s$ given 
by $4 \leq p \leq m-4$ as our criterion for fitting.
In this manner, we will only consider average over 16 or more 
samples and each subrecord will have at least 16 ranks.

\subsection{Results from literary corpora}
\label{sub:result}

\begin{figure}
\psfig{file=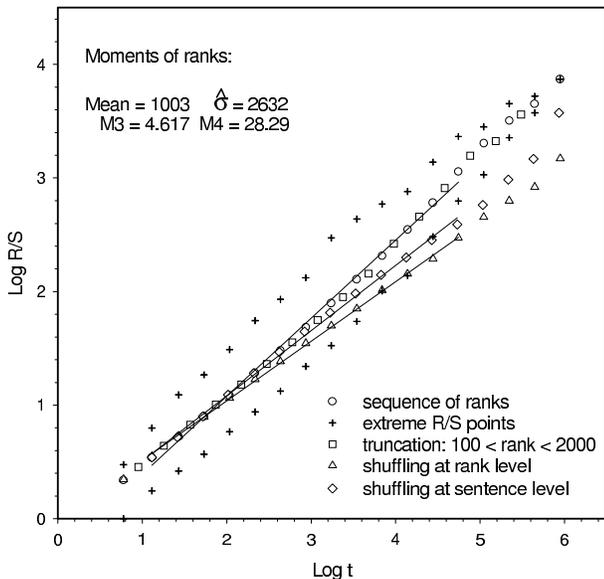,width=80mm}
\caption{R/S diagram corresponding to the Shakespeare corpus.
The linear fits are showed only for the original sequence and
the shuffling experiments.}
\label{Shake}
\end{figure}

In Fig.~\ref{Shake} we show the R/S diagram corresponding to the
coded sequence consisting in 885534 ranks from 36 plays by William
Shakespeare. The total number of different ranks, associated with
different words in the corpus, is in this case equal to 23150.
The mean and standard deviation~($\hat\sigma$) of the original 
sequence and the third (M3) and fourth (M4) moments of the 
normalised sequence are shown in an inset.
For simplicity's sake, we only mark in the diagrams the minimum 
and maximum points for a given $\log s$, and the corresponding
sample average is plotted as a small circle. The trendline is drawn
as a solid line along the points effectively used in the fitting.
The measured value of $H$ and its error are displayed in 
Table~\ref{Tfull}, and confirm the presence of long-range
correlation in the series.
In the same figure, we can also see as small squares the diagram for
the sample average from the sequence after deleting all ranks outside
the interval $(100, 2000)$. Strikingly, the corresponding value of
$H$ is statistically indistinguishable from that of the original
sequence. This fact tells us that the core of long-range correlation
is neither supported by the most frequent words nor the least used.

There are two additional experiments which can provide
information in tracking down the source of the correlated 
behaviour. The first one is a simple random shuffling of
all the ranks in the sequence which has the effect of recasting
Shakespeare's plays into a nonsensical realisation, keeping
the same original words without discernible order at any level. 
As is clear, after this experiment the analysis must yield a Hurst 
exponent indicating the uncorrelated character of the sequence. 
This is corroborated also in Fig.~\ref{Shake} where the sample 
average is plotted as small triangles.
Yet more interesting is the analysis of a shuffling of Shakespeare's
plays that preserves sentence structure, and therefore English
grammar. That is, by defining a sentence as the sequence of words
between two periods, we can reorder them in a random fashion and
thus produce a grammatically correct, though hardly meaningful,
version of the corpus. The sample averages are plotted in this
case as small diamonds. The resulting value of $H$ shows that
grammar is not sufficient to induce long-range correlations as
we can see again from Table~\ref{Tfull}. Let us note that for
maintaining the readability of the graph we only included the
extreme R/S points of the original sequence.

\widetext
\begin{table}
\begin{tabular}{lclcc}
source & original sequence & truncation
\tablenotemark[6] &
sentences shuffled & ranks shuffled \\
\hline
Shakespeare \tablenotemark[1] &
$0.687 \pm 0.040$ & $0.658 \pm 0.036$ &
$0.574 \pm 0.035$ & $0.524 \pm 0.020$ \\

Dickens \tablenotemark[2] &
$0.738 \pm 0.033$ & $0.660 \pm 0.034$ &
$0.573 \pm 0.025$ & $0.520 \pm 0.021$ \\

Darwin \tablenotemark[3] &
$0.745 \pm 0.045$ & $0.678 \pm 0.043$ &
$0.576 \pm 0.033$ & \\

Simon's model \tablenotemark[4] &
$0.550 \pm 0.040$ & $0.519 \pm 0.032$ \tablenotemark[7]& & \\

Markovian text \tablenotemark[5] &
$0.533 \pm 0.028$ & & &
\end{tabular}
\tablenotetext[1]{36 plays:  885534 words}
\tablenotetext[2]{56 books: 5616403 words}
\tablenotetext[3]{11 books: 1508483 words}
\tablenotetext[4]{$5 \times 10^6$ words generated after a transient
and deleting the ranks $\leq 5$}
\tablenotetext[5]{$1.2 \times 10^6$ words generated from table of
frequencies corresponding to the Shakespeare corpus with memory
of 7 letters}
\tablenotetext[6]{Unless other specification all ranks outside
the interval $100 < \mbox{rank} < 2000$ were deleted from
the coded sequence}
\tablenotetext[7]{In this case the ranks outside the interval
$5 < \mbox{rank} < 10000$ were deleted}
\caption{Values of $H$ from the estimation
by linear regression of $\log{\cal E}[R/S]$ vs. $\log s$.}
\label{Tfull}
\end{table}
\narrowtext

\begin{figure}
\psfig{file=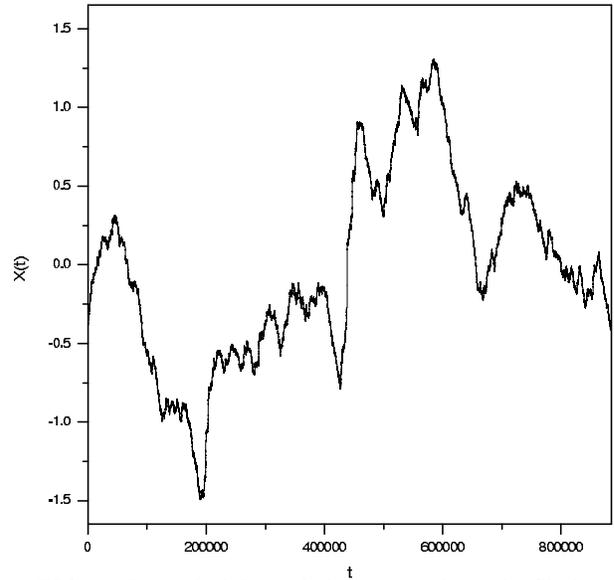,width=80mm}
\caption{Record of the coded sequence from the Shakespeare corpus.}
\label{Srecord}
\end{figure}

\begin{figure}
\psfig{file=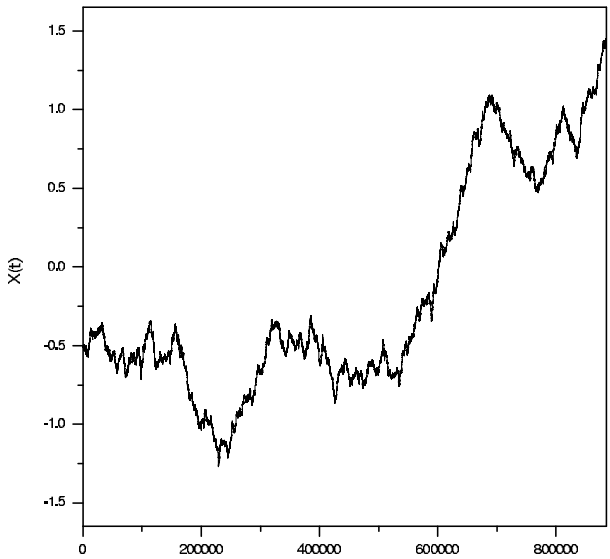,width=80mm}
\caption{Record of a fractional brownian motion generated 
with $H=0.7$}
\label{FBM}
\end{figure}

At this point it may be illustrative to compare a plot of the
record $X(t)$, the position of the walker as a function of time,
both for the Shakespeare's plays and for a stochastic sequence
generated with~$H=0.7$ by the successive random addition
method~\cite{Voss85}. As can be seen in Figs.~\ref{Srecord}
and~\ref{FBM} both records show strong persistence that manifests
itself in the long spans of average monotonous behaviour in the
records.

\vspace{2.0cm}

\newpage

\begin{figure}
\psfig{file=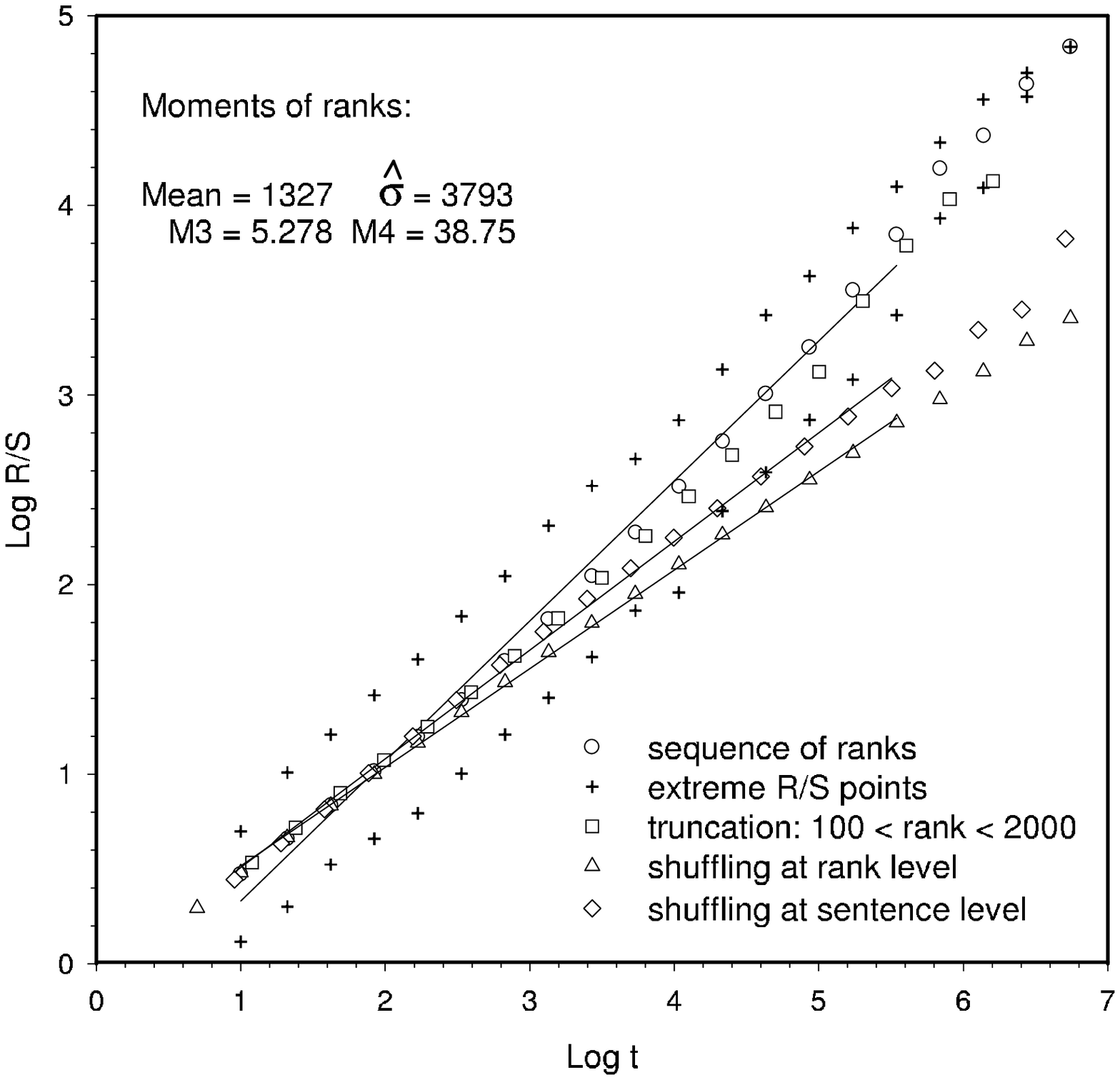,width=80mm}
\caption{R/S diagram corresponding to the Dickens corpus.
The linear fits are showed only for the original sequence and
the shuffling experiments.}
\label{Dick}
\end{figure}

\begin{figure}
\psfig{file=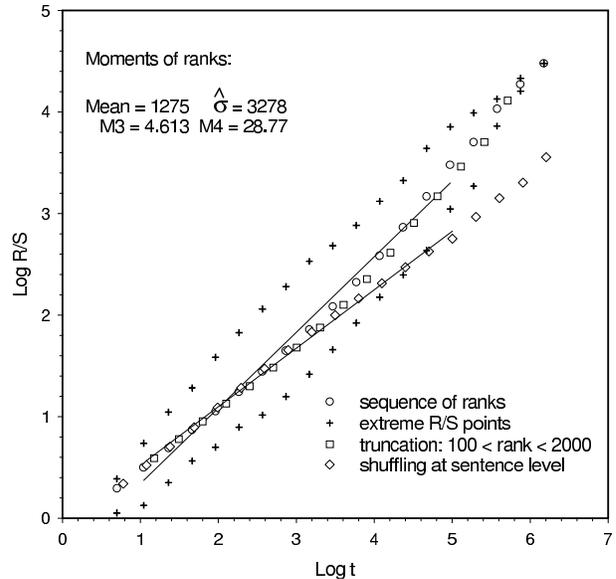,width=80mm}
\caption{R/S diagram corresponding to the Darwin corpus.
The linear fits are showed only for the original sequence and
the shuffling at sentence level.}
\label{Darw}
\end{figure}

In Figs.~\ref{Dick} and~\ref{Darw} we reproduce an identical
analysis from two text corpora gathering a collection of works 
by Dickens and Darwin respectively, whose results are summarised
in Table~\ref{Tfull}. The Dickens sequence was obtained from 56 
books by the author, and has 5616403 ranks (words) in length 
(44700 different ranks), whereas the Darwin sequence was obtained 
from 11 books and has a length of 1508483 ranks (30120 words
in the vocabulary). 
It is worth noticing that the values of $H$ from the original
sequences are indistinguishable for these two texts, written 
in prose and with different styles, although they are slightly 
greater than the value obtained for the Shakespeare sequence.
However, the three values of $H$ corresponding to the truncation
($100 < \mbox{rank} < 2000$) are statistically equivalent.
This fact suggests that the long-range correlations associated to
words in the interval considered are a robust phenomenon over
different styles and authors. Finally, the shuffling experiments 
show the same behaviour for the three authors, and are consistent 
with uncorrelated sequences.

\vspace{1.5cm}

\begin{table}
\begin{tabular}{lcc}
source       & Shakespeare       & Dickens \\
\hline
corpus       & $0.687 \pm 0.040$ & $0.738 \pm 0.033$ \\
portion      & $0.675 \pm 0.044$ & $0.715 \pm 0.043$ \\
             & $0.672 \pm 0.041$ &
\end{tabular}
\caption{Test of consistence: We estimated the $H$ value for portions
of two corpora considered in Table~\ref{Tfull}. The portions
from the Shakespeare corpus correspond to the first and second half
of the original source. From the Dickens corpus we have taken an
embedded succession of 861038 words from an arbitrary origin in the
original corpus.}
\label{Ttest}
\end{table}

\subsection{Tests of consistence}
\label{sub:cons}

In this subsection we present three tests of consistency of
our analysis. First, we performed  the same constructions as
developed in the previous subsection over portions extracted from
the original text corpora. Thus, we split the Shakespeare corpus into
two parts of equal length and measured $H$ for each part.
On the other hand, from the Dickens corpus we extracted a succession
of 861038 words from an arbitrary origin and then measured the
corresponding value of $H$. In Table~\ref{Ttest} the values from
both the original corpora and the parts are confronted.
Values indistinguishable of $H$ result from the original corpus and
its portions. In interpreting this result we should have in mind that
Zipf's analysis of a portion of the text generates a quite different
table of ranks from the one corresponding to the entire corpus.
This serves as an indication that we are quantifying a robust
phenomenon inherent to the fractal structure of texts.

In the insets of Figs.~\ref{Shake} and~\ref{Dick} we present
the values of mean and standard deviation~($\hat\sigma$) 
corresponding to the set of ranks in the entire corpus of 
Shakespeare and Dickens respectively. For the Shakespeare 
corpus we can read that the mean of ranks is equal to 1003, 
whereas $\hat\sigma = 2632$. 
On the other hand, Fig.~\ref{Dick} shows for the Dickens corpus
that the mean of ranks is equal to 1327 and $\hat\sigma = 3793$.
These quantities are calculated from sample values of a probability
distribution $P(r)$, for which Zipf's law represents a crude 
approximation. However, by using more accurate analytical
descriptions of the probability density $P(r)$, it is possible 
to evaluate the statistical moments for the whole distribution,
that is, in the limit of infinite vocabulary.
In particular, let us mention that the standard deviation 
calculated for both the Shakespeare and Dickens corpora, using the
analytical expressions for $P(r)$ obtained in Ref.~\cite{Monte01},
result finite in the infinite vocabulary limit.
The reason for this is that for large values of $r$, the probability
density develops a fast exponential decay in the case of
single-author corpora~\cite{Monte01}.

In Table~\ref{Tfull} we also report the values of $H$ corresponding
to corpora generated by means of stochastic processes. 
Particularly, we did the analysis explained above over a sequence 
of ranks generated by the Simon's model~\cite{Simon} and other 
corresponding to a Markovian text~\cite{Hayes}, with a memory of
seven letters. This short-range memory is enough to string out 
groups of few words in grammatical order.
However, as we can see from the values on the table, we do not
obtain correlations from neither type of sequence, which is
consistent with the expected behaviour for processes with
short-range correlations. Consequently, our implementation of
the {\em rescaled range} analysis allows to clearly distinguish
between real and stochastic version of texts~\cite{CMH97}.

\subsection{The Dictionary}
\label{sub:Webster}

So far, we still do not have enough evidence as to assert a 
precise source for the phenomenon being analysed in this paper.
We have characterised with a robust quantifier the presence of
long-range correlations in literary texts beyond sentence level,
and in fact over ranges spanning more than one individual work.
The shuffling experiments attest where the long-range correlation 
does not originate, but say nothing on where it does.
In previous attempts to study this phenomenon, a variety of possible
origins for the long-range dependence have been put forward, though
the conclusions in all cases have been inferred from the observed
statistical behaviour of single literary works, and from the mapping
of texts at letter level.
In the present work we use a robust mapping with a strong footing on
linguistics, and focus on the correlations that arise in large corpora
comprising many individual books with no thematic linkage. 
Therefore, the existence of correlations overarching sets of entire
works should be attributed to something else than either the relation
between ideas expressed by the author as it was proposed in
Ref.~\onlinecite{SZZ93}, or nonuniformities in the distribution of
word's lengths and the associated densities of blank spaces as it
was suggested in Ref.~\onlinecite{Ebe}. In particular, the latter
alternative is ruled out from the outset since our mapping does not
carry information on word structure.

\begin{figure}
\psfig{file=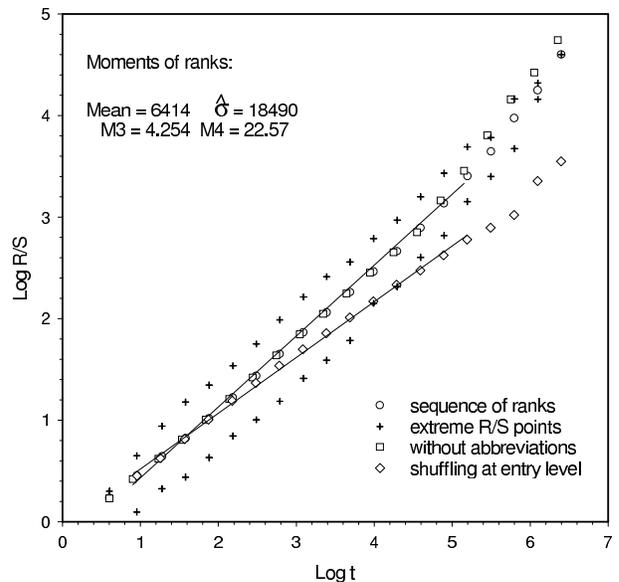,width=80mm}
\caption{R/S diagram of the Webster's abridged dictionary
(1913 edition). The linear fits correspond to the sequences 
without the abbreviations.}
\label{Webs}
\end{figure}

In order to gain more insight into the origin of the attested
long-range correlations, we decided to perform another experiment 
with a special type of text. A dictionary is a large collection of
entries, one for each different word, arranged in alphabetical order.
Clearly, in general, sequences of thematic affinity span at most 
a few entries, for small groups of words associated by common roots.
Nevertheless, as is confirmed in Fig.~\ref{Webs} where we display the
R/S diagram generated from the Webster's abridged dictionary (1913
edition), the value of $H$ certifies the presence of long-range
correlations. Table~\ref{Twebs} compiles the exponent $H$ from the
entire dictionary and from the text without the abbreviations (which
indicate the grammatical category of entry words), obtaining in both
cases totally consistent values for $H$.
This result is rather striking and hints at the presence of a 
type of long-range organisation in the layout of lexical entries 
that may be connected with the phenomenon observed in literary works. 
This last assertion is further supported by performing a shuffle
over the whole dictionary keeping integrity at entry level.
That is, we do not disrupt the structure of entries made up of each
word together with the associated definition, but we do mix them
in a random manner. The resulting collection of definitions
now lacks the alphabetical ordering; however, the explicit
information contained in the dictionary is still intact.
Notwithstanding, the long-range order is completely obliterated by
the shuffle as accounted for by the fall in the $H$ value down to
that of an uncorrelated sequence.
This experiment corroborates that beyond the local structures derived
from semantic affinity of small groups of related words, the ordered
lexicon also possesses an overall macrostructure that emerges, as a
hidden layer, out of the alphabetical order of entries.
 
A close examination at the structure of the dictionary reveals the
source of the long-range order. Let us take for example the following
sequence of related entries from the Webster's abridged dictionary
(1913 edition): {\em advice, advisability, advisable, advisableness,
advisably, advise, advised, advising, advisedly, advisedness,
advisement, adviser, advisership, advisor}, and {\em advisory}.
These entries form a small cohesive block of information, and 
all their definitions share many words. In turn, these shared terms 
point to, possibly, remote locations in the dictionary where, again, 
words with common roots form small clusters. This process carries on 
to many levels of depth, thus building a complex network of relations 
at word level, which may be closely related to the presence of 
long-range correlation in the succession of words in the dictionary.  
The entry--level shuffling destroys that emerging order, and thereby
the correlations. 

\vspace{1.0cm}

\begin{table}
\begin{tabular}{lc}
Webster's Dictionary & $H$ \\
\hline
original sequence      & $0.690 \pm 0.031$ \\
sequence without abbr. & $0.699 \pm 0.036$ \\
entries shuffled
\tablenotemark[1]      & $0.548 \pm 0.025$
\end{tabular}
\tablenotetext[1]{from the text without abbreviations}
\caption{Values of $H$ for the Webster's abridged dictionary
(edition 1913) after coding as sequence of ranks.}
\label{Twebs}
\end{table}

\section{Conclusions}
\label{sec:fin}

In this work we have addressed the important issue of the
emergence of long-range correlations in human written
communication.  To that end we proposed a simple mapping of texts
onto random walks that keeps {\em the word} as the basic unit of
communication. Therefore, the time series generated by this mapping
retain the structure relevant to the linguistic phenomenon being
analysed.
 
We addressed in some detail the {\em rescaled range} analysis,
which has been successfully applied to the analysis of a vast 
variety of time series exhibiting long-term correlations.  
By applying this analysis to the coded texts we found conclusive
evidence for long-range correlation in the use of words over 
spans  as long as the whole corpora under study. It is worth 
noticing  that the corpora used in this work are made up of the 
concatenation of independent literary works by individual authors, 
and, therefore the long-range effects must emerge as a phenomenon
independent of the particular bounds of single literary works.
This observation led us to analyse the case of a dictionary as a
special kind of text, which provided insight into possible sources
for the long-range order.  The alphabetical order of entries in a
dictionary corresponds only to the first visible layer of structural
organisation.  Yet, by performing the shuffling at entry level we
realised that the whole of the long-range order is dependent upon a
more complex structural layer related to the network of associations
among word clusters.

As it has been convincingly shown in Ref.~\cite{MZ01} sets of
literary works also possess higher level structures associated with
systematic patterns in word usage, which might also give rise to 
a complex  network topology of relations among groups of words. 
The detailed mechanisms whereby the onset of these structures takes 
place in language requires further interdisciplinary research. 
In the light of the evidence supplied in this work, it is 
plausible that the ultimate source of these correlations is 
deeply related to the structural patterns of word distribution 
in written communication.
In this view, groups of words associated by affine semantic
hierarchies form a complex arrangement of cohesive clusters
even over spans of entire corpora, thereby giving rise to
long-range order in human written records.

\acknowledgments
The authors are grateful to D.~H.~Zanette for his
critical reading of the manuscript and to
F.~A.~Tamarit for useful suggestions.
This work has been partially supported by
``Secretar{\'\i}a de Ciencia y Tecnolog{\'\i}a de la Universidad
Nacional de C\'ordoba'' (fellowship and grant 194/00).
The digital texts analysed in this paper were obtained from
{\em Project Gutenberg Etext}~\cite{Gut}

%

\end{document}